\documentclass[a4paper]{toledoStyle}
\usepackage{epsfig}

\begin{document}

\title{Fragmentation in Kinematically Cold Disks}

\author{D. Huber \and D. Pfenniger} 

\institute{
  Geneva Observatory, CH-1290 Sauverny, Switzerland}

\maketitle 

\begin{abstract}
Gravity is scale free. Thus gravity may form similar structures in
self-gravitating systems on different scales. Indeed, observations
of the interstellar medium, spiral disks and cosmic structures, 
reveal similar characteristics. The structures in these systems
are very lumpy and inhomogeneous. Moreover some of these 
structures do not seem to be of random nature, but obey 
certain power laws. 

Models of slightly dissipative self-gravitating
disks show how such inhomogeneous structures can be maintained 
on the kpc-scale. The basic physical processes 
in these models are self-gravity, dissipation and differential 
rotation. In order to explore the structures resulting from these 
processes, local simulations of self-gravitating 
disks are performed in 2D and 3D.  
We observe persistent 
patterns, formed by transient structures, whose intensity 
and morphological characteristic depend on the dissipation rate.

\end{abstract}

\section{Introduction}
  
Molecular clouds reveal hierarchical structures, obeying power laws
over several orders of magnitude in scale. Observations suggest, that 
the hierarchical structure of kinematically cold media is not only 
present in Milky Way molecular clouds, but is also found in other 
systems and on larger scales. \cite*{authorf:Vogelaar94} found, 
e.g., perimeter-area correlations in high-velocity clouds. 
Power-law power spectra were found in the Small and the Large 
Magellanic Cloud by \cite*{authorf:Stanimirovic99} 
and \cite*{authorf:Elmegreen01}, respectively. 
Furthermore, measurements of the HI distribution 
in galaxies of the M81 cluster reveal fractal structures on 
the galaxy disc scale (\cite{authorf:Westpfahl99}). 
The matter on cosmic scales is also hierarchically organized. 
A common feature of the ISM and the cosmic structure is, that the 
matter distribution can be characterized by a comparable fractal dimension.
All this suggests, that a general scale free factor is mainly 
responsible for the matter distribution and the dynamics of cosmic 
structures, disks and molecular clouds. There is only one factor 
being able to have a dominant influence on all these scales, namely
gravity. The local shearing sheet experiments of 
Toomre \& Kalnajs 
(1991, hereafter TK) show that gravity in combination with shearing
and dissipation can develop long-range correlations and maintain 
the system out of equilibrium. In order to study in detail the
structures resulting from these processes on the kpc scale, we 
perform local simulations of self-gravitating disks in 2D and 3D.
The third dimension becomes important as soon as a strong matter
clumping causes a tight coupling of self-gravitating forces in 
the 3D equations of motion.

\section{Local Model}

\subsection{Principle}

In local models of disks, everything inside a box of a given size
is simulated and more distant regions in the plane
are represented by replicas of the local box. In such a model the 
orbital motion of the particles is determined by Hill's approximation
of Newton's equation of motion (\cite{authorf:Hill78}). In 3D they read
\begin{equation}
\begin{array}{ccccccc}
\ddot{x}&-&2\Omega_0\,\dot{y}&=&4\Omega_0 A_0 x& + &F_x(x,y,z)\;\;\\
\ddot{y}&+&2\Omega_0\,\dot{x}&=& & &F_y(x,y,z)\;\;\\
\ddot{z}& &                  &=&-\nu^2 z&+&F_z(x,y,z)\;,
\end{array}
\label{authorf_eq:eq1}
\end{equation}
where $A_0=-{1\over2} {\cal R}_0(d\Omega /d{\cal R})_{{\cal R}_0}$ 
is the Oort constant of
differential rotation and $\nu$ is the vertical epicycle frequency.
$F_x$, $F_y$ and $F_z$ are local forces due to 
self-gravitating particles.

The local system is periodic in the $y$-direction and isolated in 
the $z$-direction. 
If we use an affine coordinate system whose pitch angle changes with 
time then the system is also in the $y$-direction periodic 
(\cite{authorf:Huber99}, 2001). Thus we can calculate the 
forces with the convolution method using the FFT algorithm. 
Thereby the computation time is reduced to be proportional 
to $N_{\rm c}\log(N_{\rm c})$, where $N_{\rm c}$ is the number 
of cells, taken here as proportional to the number of particles. 

\begin{figure*}[!t]
\centerline{
\epsfig{file=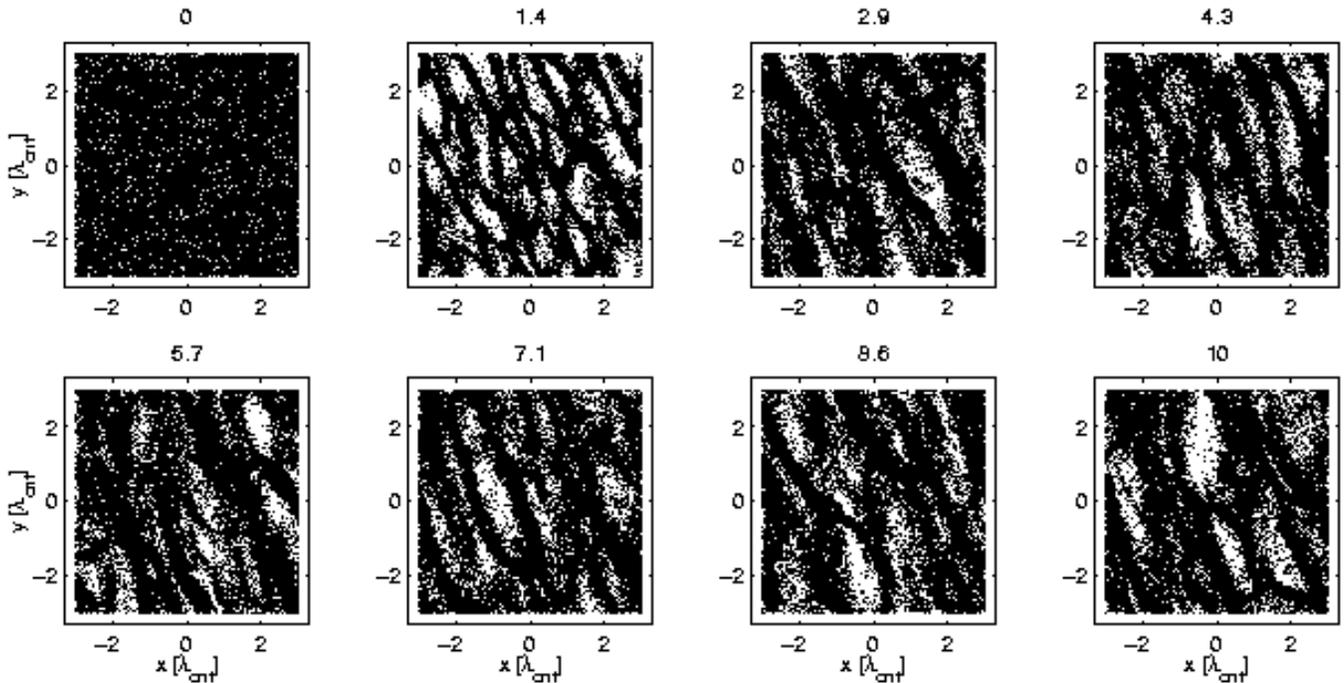,angle=90,width=17.72cm}}
\caption{The evolution of the particle positions seen from above the
  galaxy plane. The structures result from a 3D simulation with a ``weak'' 
  dissipation.}
  \label{authorf_fig:huberd1_fig1}
\end{figure*}

\begin{figure*}[!t]
\centerline{
\epsfig{file=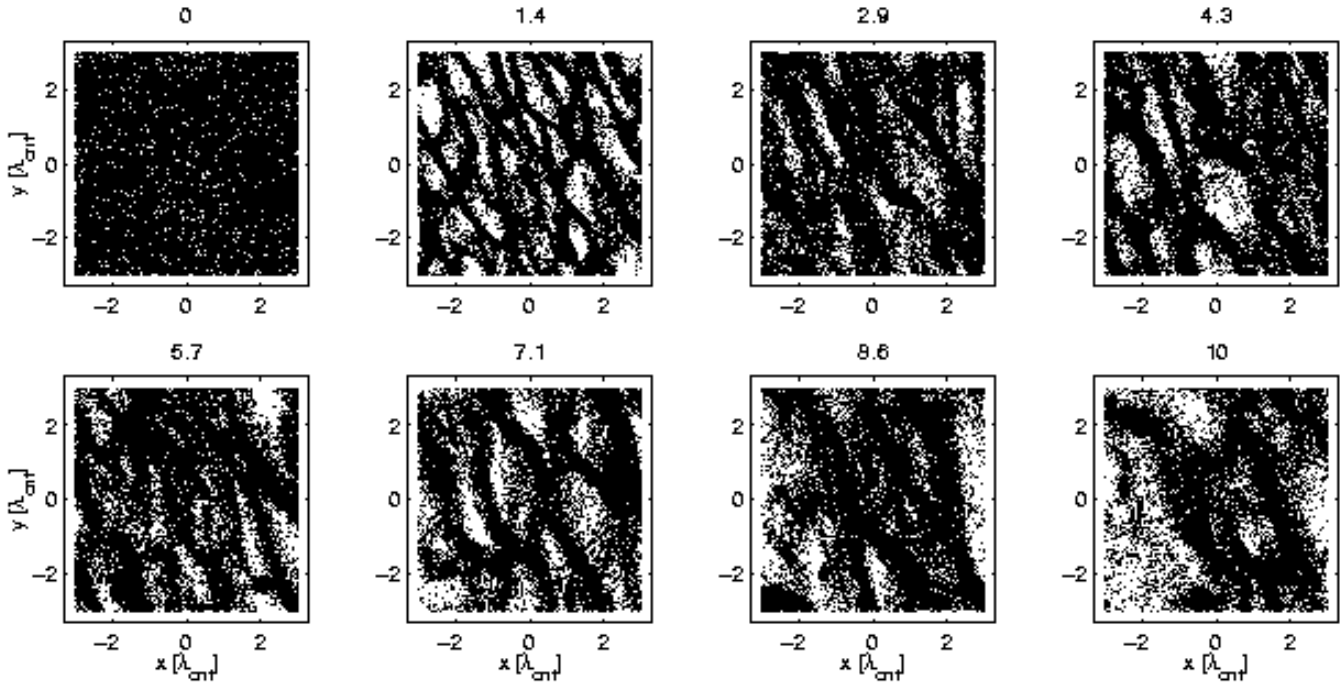,angle=90,width=17.72cm}}
\caption{The evolution of the particle position, resulting from a
  simulation with a ``middle'' dissipation.}
  \label{authorf_fig:huberd1_fig2}
\end{figure*}

\begin{figure*}[!t]
\centerline{
\epsfig{file=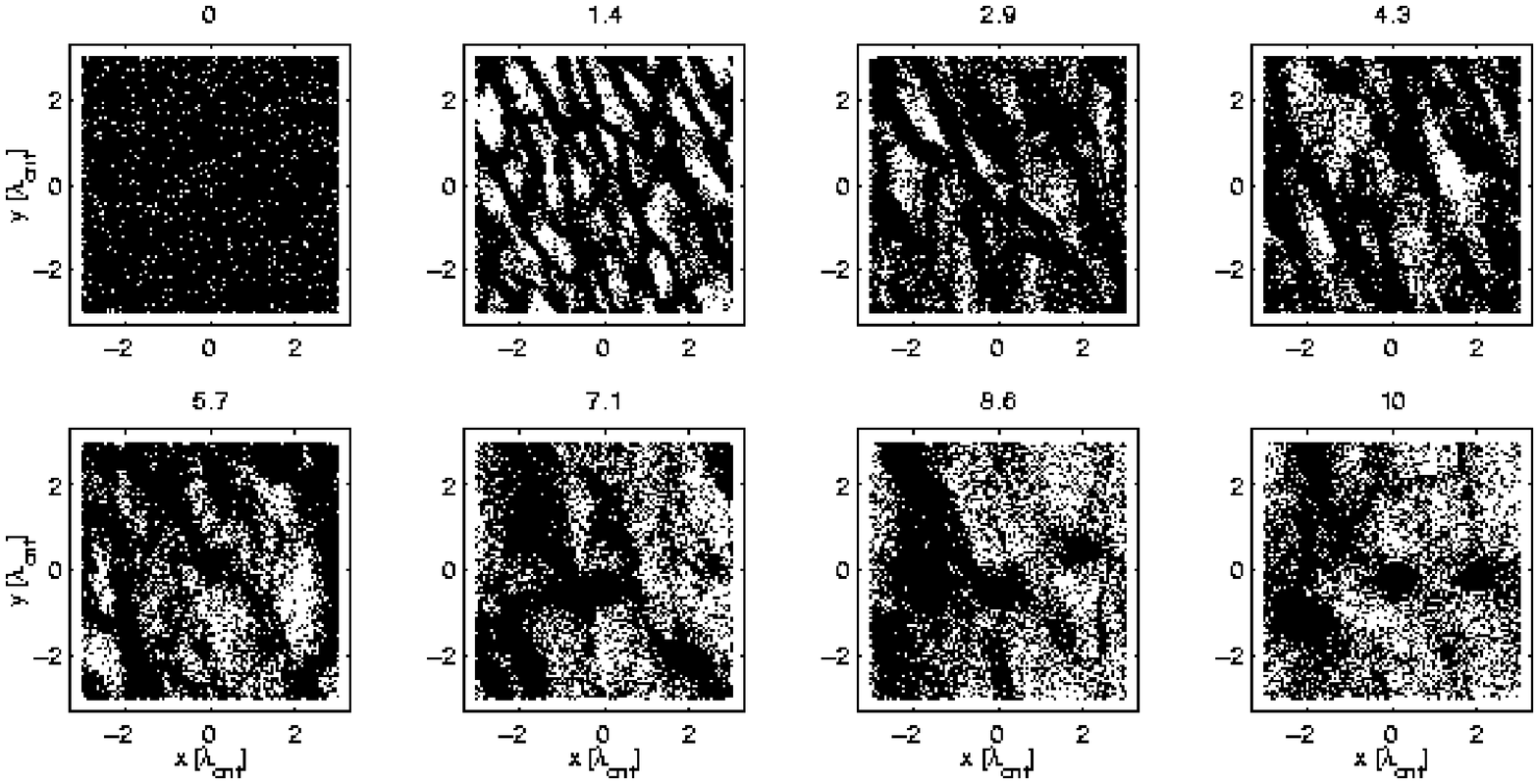,angle=90,width=17.72cm}}
\caption{The evolution of the particle position, resulting from a
  simulation with a ``strong'' dissipation.}
  \label{authorf_fig:huberd1_fig3}
\end{figure*}

\subsection{Weak Friction}

To counteract the particle dynamical heating, TK proposed to add 
an ad-hoc friction term playing the role of the dissipative factors 
at work in the interstellar gas. Following TK we include linear
friction terms as well. The friction terms should be weak in
order to keep a quasi-Hamiltonian system. Indeed at the kpc scale the
physics is still dominated by gravitational dynamics and its concrete
behavior should be weakly dependent on the particular dissipative factors. 

The linear friction terms $-C_x\dot{x}$ and $-C_z\dot{z}$
added to the radial resp. vertical forces $(F_x, F_z)$ control 
the particle motions via Equation~\ref{authorf_eq:eq1}. 
There is no azimuthal friction in
order to be
consistent with a global angular momentum conservation.

\subsection{Initial Conditions}

We are interested in the secular time behavior of the galaxy disk.
Thus we perform simulations for $t=10$ galactic rotations.
In the initial state at $t=0$
the particles are distributed uniformly in the $x$-$y$-plane. In the
$z$-direction the particle distribution follows an isothermal law
\begin{equation}
\rho\propto\mathop{\rm sech}{}^2(z/z_0)\, , 
\end{equation}
where $\rho$ is the density and $z_0$ is the disk scale height.
The velocities at $t=0$ are determined by the shear
\begin{equation}
\dot{x}=0\:,\;\;\;\dot{y}=-2A_0 x\:,\;\;\; \dot{z}=0
\end{equation}
and the Schwarzschild velocity ellipsoid
\begin{eqnarray}
\sigma_x &=&\frac{3.36 G\Sigma_0 Q}{\kappa}\nonumber\\
\sigma_y &=&\frac{\sigma_x\kappa}{2 \Omega_0}\\
\sigma_z &=&\sqrt{\pi G \Sigma_0 z_0}\;, \nonumber 
\end{eqnarray}
where the Safronov-Toomre stability criterion is $Q \geq 1$.

\section{Results}

We present here some preliminary results of an extended study to this
subject (\cite{authorf:Huber01}).

\subsection{3D shearing boxes}

The structures resulting from shearing box simulations depend a
lot on the relative strength of the competing gravitational and 
dissipation processes. Gravitational instabilities lead to a 
conversion of directed kinetic energy (shear-flow) into random 
thermal motion. In this way the disk is heated up. If the dissipation 
is too weak then initially arised structures can not be maintained 
and smear out quickly. If the dissipation is increased, a filamentary 
structure can be maintained in a statistical equilibrium. If we 
continue to increase the dissipation the filaments become denser and
denser and clumps in filaments may be formed. If finally the dissipation 
dominates completely the heating process hot clumps, collecting almost 
all the matter of the simulation zone are formed out of the
filaments. The change of the structure morphology for an 
increasing dissipation is showed in Figures~1-3.
For convenience we call the dissipation strengths, leading to the
presented structures, ``weak'', ``middle''and ``strong''.  
The relative cooling times corresponding to these dissipation
strengths are:  
$\tau_{\rm cool,1} : \tau_{\rm cool,2} : \tau_{\rm cool,3} : 
\tau_{\rm osc} = 16 : 15 : 14 : 1$, 
where $\tau_{\rm osc}$ is the period of the unforced epicyclic motion.

The simulations are carried out
with 131040 particles, corresponding to a surface 
number density of
$n=3640\:\lambda^2_{\rm crit}$. The dynamical range is $1.8$ dex.
Only each forth particle is shown.

\subsection{Mass-Size Relation}

In order to characterize the structures resulting from the shearing
box experiments, we determine the mass-size relation. 
We choose a representative set of particles and count for each
particle the number of neighboring particles $N(R)$ inside a certain
radius. If we repeat this for other values of $R$ we can find
the structure dimension $D(R)$ via
\begin{equation}
D(R)=\frac{d {\rm ln}(N)}{d{\rm ln}(R)}(R)\;,
\end{equation}
where $R$ denotes the scale.
The mass-size relation is then
\begin{equation}
N(R)\propto M(R) \propto R^{D(R)}\;.
\end{equation}
If the structure dimension is independent of the scale, $D=D_f$, 
i.e., if $D$ is constant or oscillates around a mean value, then the 
mass-size relation is a power law, 
\begin{equation}
M \propto R^{D_f}\;.
\end{equation}
If furthermore $D_f$ is non-integer, the structure is fractal.
If however the structure dimension depends on the scale $D=D(R)$,
the structure dimension may simply be regarded as a statistical
measure describing the clumpiness at the corresponding scale.

In order to asses the general relevance of the underlying physical
processes we check the structure for self-similarity, i.e., we check
if $D=D_f$ for a certain scale range. However, one has to take into
account that the structures result form a finite simulation, 
modeling a finite physical system. Thus the system can not be fractal
beyond an upper and lower cutoff. An upper cutoff is given by the
scale at which the system become periodic. 
A lower scale limit is due to the finite resolution of
the simulation mesh.

\begin{figure}[!ht]
\centerline{
\epsfig{file=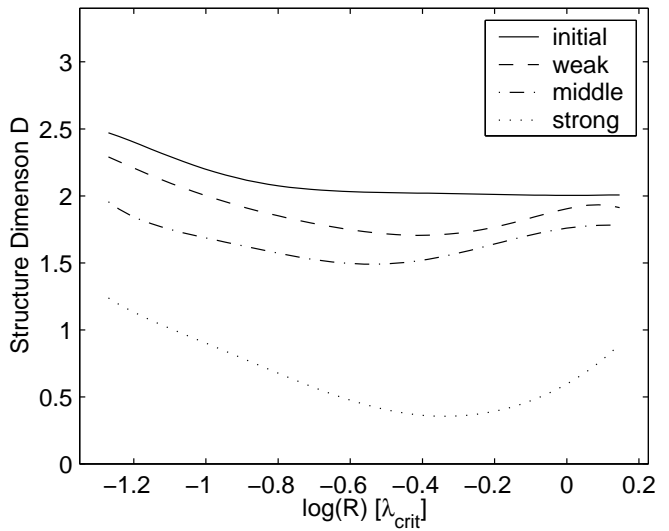,angle=90,width=\hsize}}
\caption{The structure dimension $D$ of the mass distributions showed
  in Figures~1-3 as a function of the scale $R$. The corresponding
  simulations were carried out with a ``weak'', a ``middle'' and a 
  ``strong'' dissipation, respectively. The solid line depicts the
  initial mass distribution. On large scales this state represents a
  2D matter distribution, whereas on small scales it tends to $D=3$.}
  \label{authorf_fig:huberd1_fig4}
\end{figure}

Figure~\ref{authorf_fig:huberd1_fig4}  shows the structure dimension of the 
mass distribution presented in Figures~1-3. The longer term evolution
of the structure may be superimposed by
fluctuations on time-scales of the order of
$\sim 1/2\; \tau_{\rm rot}$, where
$\tau_{\rm rot}$ is the time for a rotation around the galaxy center.
In order to eliminate these fluctuations we indicate in this paper 
mean values of the structure dimension $D$ determined 
during the last two rotations.

The structure dimensions are not constant over the corresponding 
dynamical range and are thus not fractal. However the structure 
dimension resulting from the
simulation with the ``middle'' dissipation has a 
structure dimension $1.5<D<1.8$ over the whole dynamical range
and remains smaller than 2 also on scales where the disk
thickness becomes important. Thus the corresponding matter 
distribution can approximately be described by a power-law
for the considered scale range. The structure dimension
has a minimum at $R=0.25\:\lambda_{\rm crit}$. The increase
of the structure dimension at $R>\mid 0.25\:\lambda_{\rm crit}\mid$ 
may then be due to the lower and upper cutoff. 
A larger dynamical range may
thus flatten the curve depicting the structure dimension. This
supposition is supported by low resolution simulations showing
a steeper increase of $D$ beyond the minimum.

\section{Conclusions}

The structure resulting from the local simulations of 
self-gravitating disks can be homogeneous, filamentary or 
clumpy depending on the relative strength of the competing 
gravitational and dissipation processes. As long as the structure 
is mainly filamentary self-gravitation and dissipation ensure a 
statistical equilibrium, i.e., persistent patterns consisting of 
transient structures are formed. If the dissipative processes 
begin to dominate the evolution, the structures turn from 
filamentary to clumpy. During the subsequent evolution the 
clumps become hotter and more massive. 
In general clumpy structures do not evolve towards a 
statistical equilibrium. However 2D simulations with a dynamical 
range of 2.5 dex  show that it is also possible to 
establish persistent patterns formed by clumps.

A larger dynamical range produce in general a flatter structure
dimension curve. However the scale range of the 
simulations is still too small to draw final conclusions about 
self-similarity in open, self-gravitating systems.

\begin{acknowledgements}

\mbox{This work has been supported by the Swiss Science Foundation.}

\end{acknowledgements}

\end{document}